\documentclass[twocolumn,english,aps,prb,floatfix,preprintnumbers,showpacs,amsfonts,amssymb,superscriptaddress]{revtex4}
\usepackage{ae,aecompl}
\usepackage[T1]{fontenc}
\usepackage[latin1]{inputenc}
\usepackage{amsmath}
\usepackage{graphicx}
\usepackage{amssymb}

\makeatletter
\usepackage{amscd}
\usepackage{bm}

\usepackage{babel}
\makeatother
\usepackage{color}

\begin{document}

\title{Coexistence of Pairing Tendencies and Ferromagnetism \\
in a Doped Two-Orbital Hubbard Model on Two-Leg Ladders}

\author{J.~C.~Xavier}

\affiliation{Instituto de F\'{\i}sica, Universidade Federal de Uberl\^andia, Caixa
Postal 593, 38400-902 Uberl\^andia, MG, Brazil}

\author{G.~Alvarez}

\affiliation{Computer Science \& Mathematics Division and Center for Nanophase
Materials Sciences, Oak Ridge National Laboratory, Oak Ridge, Tennessee
37831, USA}

\author{A.~Moreo}

\affiliation{Department of Physics, University of Tennessee,
Knoxville, TN 37996 and
Materials Science and Technology Division, Oak Ridge National Laboratory,
Oak Ridge,TN 37831} 

\author{E.~Dagotto}

\affiliation{Department of Physics, University of Tennessee,
Knoxville, TN 37996 and
Materials Science and Technology Division, Oak Ridge National Laboratory,
Oak Ridge,TN 37831}

\date{\today{}}

\begin{abstract}
Using the Density Matrix Renormalization Group and two-leg ladders, 
we investigate an electronic two-orbital Hubbard model including plaquette
diagonal hopping amplitudes.
Our goal is to  search for regimes where
charges added to the undoped state form pairs, presumably a precursor of
a superconducting state.
For the electronic density $\rho=2$, i.e. the undoped limit, 
our investigations show a robust $(\pi,0)$ antiferromagnetic ground state,
as in previous investigations. 
Doping away from $\rho=2$ and for large values
of the Hund coupling $J$, a ferromagnetic region is found to be stable. 
Moreover, when the interorbital on-site Hubbard repulsion is smaller than
the Hund coupling, i.e. for $U'<J$ 
in the standard notation of multiorbital Hubbard
models,  our results indicate the coexistence
of pairing tendencies and ferromagnetism close to $\rho=2$. These results are
compatible with previous investigations using one dimensional systems.
Although further research is needed to clarify 
if the range of couplings used here is of
relevance for real materials, such as superconducting heavy fermions or pnictides, 
our theoretical results address a possible mechanism for pairing that
may be active in the presence of short-range ferromagnetic fluctuations.
\end{abstract}


\maketitle

\section{INTRODUCTION}

It is widely believed that magnetism is a fundamental ingredient 
to explain the origin of high-temperature superconductivity
in several materials. In fact, there is experimental evidence that the superconductivity
in many heavy fermion (HF)  compounds is mediated by spin fluctuations.\citep{mathuretal,jourdan,revsigrist,xavierprl2008}
Mechanisms for superconductivity based on antiferromagnetism have been extensively discussed for the 
Cu-based high-temperature superconductors as well.\citep{dagottorev}
Recently, considerable excitement has been generated by the discovery
of high-temperature superconductivity in the iron pnictides.\citep{oxypnictides} 
Except for the cuprates, the iron-based superconductors now have 
the highest superconducting (SC)
critical temperature $T_{\rm c}$ of any material
(see for example Ref. \onlinecite{ironhightc}).
As in HF systems and cuprate superconductors, in the pnictides there is also evidence that the
superconductivity is not mediated by the electron-phonon interaction.\citep{ironbcs1,ironbcs2} 


Magnetism and superconductivity can appear in different ways. In some
HF compounds, superconductivity and antiferromagnetic (AFM) order
co-exist,\citep{mathuretal} while for the cuprates the
 superconductivity emerges after the long-range AFM order is destroyed by doping.\citep{dagottorev}
In some HF systems, it is the superconductivity and ferromagnetism (FM) 
(as opposed to AFM order) that
co-exist.\citep{natferro} In this work,  we will
be interested in detecting a clear evidence of pairing of extra charges that are added
to the undoped limit where magnetic order exists. Using a two-orbital model
and two-leg ladders,
it will be shown that
a possible region for the robust coexistence of (spin triplet) 
pairing together with
magnetic order is where ferromagnetism develops. This is in 
qualitative agreement
with previous investigations carried out using one-dimensional 
systems.\citep{1D}
Our effort should be 
considered simply as providing
 the first steps in relating pairing and magnetism 
in a complex two-orbital model via computational 
techniques on ladder geometries. 
Antiferromagnetic order, as found in the pnictides, could
also be favorable for pairing tendencies in the spin singlet channel,
as discussed recently.\citep{prl2orbdagotto,prb2orbdagotto}

In principle, a theoretical investigation based on model Hamiltonians 
for strongly correlated materials
starts with an effective tight-binding model, containing the minimum 
ingredients to describe the physics of the materials under investigation. 
However, even if a well-defined reasonable model is used, 
it is still highly non-trivial to extract the ground state properties of these 
effective models in 
two or three dimensions using unbiased numerical 
methods. In fact, at present there are no 
accurate techniques to study Hubbard-like 
models in dimensions two and three. Thus, in order to get at least 
some insight  on the ground
states properties of theses models, it is common practice
to study the model Hamiltonians in quasi-one-dimensional geometries. In
particular, a very popular route that has been used for
several theoretical investigations is to study strongly correlated electronic
systems using ``ladder'' geometries.\cite{scielbio}
The $N$-leg ladders consist of $N$  chains of length $L$ coupled by some
parameter (as, for example, fermionic hopping terms). 
The two-dimensional system can in principle be obtained 
by considering the limits of both $N$ and $L$ sent to
infinity, although in practice this is difficult to do. 
This ladder-based procedure has been used to investigate models for 
the high temperature superconductors \cite{ladders} and for the 
HF systems. \cite{xavier2,xavierprl2008}
Some important results were obtained with this method.
For example, research based on microscopic models
for the high $T_{c}$ superconductors,\citep{dagottorev} as well as research on
HF models,\citep{xavierprl2008} indicate that superconductivity 
mediated by antiferromagnetic  fluctuations can be stabilized, in agreement with
several experiments.  Thus, the use of ladders appears to be an
important ingredient to unveil dominant ground state tendencies. Moreover,
the hopping amplitudes that will be used in our investigations below
include next-nearest-neighbor diagonal hoppings that are only possible
when plaquettes exist in the lattice under consideration.


Note that microscopic models that may present superconductivity induced by
antiferromagnetism, such as the one-orbital Hubbard model and the Kondo Lattice model,
have been extensively studied by several authors.
However, microscopic models for \emph{superconductivity
in a ferromagnetic spin background}
have been much less explored, with the exception of studies using
one-dimensional chains.\citep{1D} 
This may be caused in part by the 
perception that superconductivity and
ferromagnetism, as opposed to antiferromagnetism, 
can not coexist.\cite{notcoexit} However, this perception has been
challenged by the discovery  of superconductivity and 
FM in the HF compounds UGe$_2$\cite{natferro} and
URhGe.\cite{natoferro2} Moreover,
SC and FM were also observed\cite{natoferro3} in the $d$-band metal ZrZn$_2$.

Motivated by the discovery of superconductivity in the HF compound
UGe$_2$, a few years ago Karchev \emph{et al.} proposed
a one-band model to study the coexistence of superconductivity and
ferromagnetism.\citep{bedell} However, other researchers have argued
that the treatment used to investigate that model, as well as the model
itself, were not appropriate to describe the coexistence of these 
phases.\citep{replybedell1,replybedell2,replybedell3}

Due to the lack of studies of microscopic models for
superconductivity mediated by ferromagnetic fluctuations beyond one-dimensional
chains,\citep{1D} in this work
we have decided to investigate a microscopic ladder model where the 
coexistence of superconductivity 
and ferromagnetism appears possible. In fact, it will be
shown below that the two-orbital model that has been 
originally proposed to describe
the low-energy physics of the iron-based 
superconductors\citep{2orbscalapino,prl2orbdagotto,prb2orbdagotto} actually
leads to the  coexistence of pairing and spin ferromagnetic  tendencies.
Qualitative our results are compatible with those reported using
chains.\citep{1D}
Although the model considered here may not
be a proper effective model for superconducting ferromagnets such as UGe$_2$, 
we believe that the
mechanism that bounds together the charges carriers (see section III.B) is so
simple and generic that our calculations
may also apply in a variety of other models as well. 

\section{MODEL}

In these studies, we have considered the following  Hamiltonian 
defined on a two-leg ladder geometry
\begin{eqnarray}
H =& - &\sum_{j,\sigma,\gamma,\gamma',\lambda,\lambda'}t_{\gamma,\gamma'}^{\lambda,\lambda'}
\left(d_{j,\gamma\sigma,\lambda}^{\dagger}d_{j+1,\gamma'\sigma,\lambda'}^{\phantom{\dagger}}+\mathrm{H.}\,\mathrm{c.}\right)\nonumber\\
&- & 
\sum_{j,\sigma,\gamma,\gamma'}\tilde{t}_{\gamma,\gamma'}\left(d_{j,\gamma\sigma,1}^{\dagger}d_{j,\gamma'\sigma,2}^{\phantom{\dagger}}+
\mathrm{H.}\,\mathrm{c.}\right)\nonumber\\
& +&
U\sum_{j,\gamma,\lambda}\mathbf{\rho}_{j,\gamma\uparrow,\lambda}\mathbf{\rho}_{j,\gamma\downarrow,\lambda}+
(U'-J/2)\sum_{j,\lambda}\mathbf{\rho}_{j,x,\lambda}\mathbf{\rho}_{j,y,\lambda}\nonumber\\
&-& 
2J\sum_{j,\lambda}{\mathbf{S}}_{j,x,\lambda}\cdot{\mathbf{S}}_{j,y,\lambda}\nonumber\\
&+&
J\sum_{j,\lambda}(d_{j,x\uparrow,\lambda}^{\dagger}d_{j,x\downarrow,\lambda}^{\dagger}d_{j,y\downarrow,\lambda}d_{j,y\uparrow,\lambda}+H.c),
\end{eqnarray}
where $d_{j,\gamma\sigma,\lambda}^{\dagger}$ creates an electron
with spin projection $\sigma$ in the orbital $\gamma=x,y$ ($d_{xz}$ and
$d_{yz}$, respectively) at the rung $j$ and leg $\lambda=1,2$,
 $ {\mathbf{S}}_{j,\gamma,\lambda}$ is the electron spin density  operator, 
$\mathbf{\rho}_{j,\gamma\sigma,\lambda}=d_{j,\gamma\sigma,\lambda}^{\dagger}d_{j,\gamma\sigma,\lambda}$,
and $\mathbf{\rho}_{j,\gamma,\lambda}=\sum_\sigma d_{j,\gamma\sigma,\lambda}^{\dagger}d_{j,\gamma\sigma,\lambda}$.

The hopping amplitudes are: $t_{x,x}^{\lambda,\lambda}=\tilde{t}_{y,y}=-t_{1}$,
$t_{y,y}^{\lambda,\lambda}=\tilde{t}_{x,x}=-t_{2}$, $t_{\gamma,\gamma}^{1,2}=t_{\gamma,\gamma}^{2,1}=-t_{3}$,
$t_{\gamma,\gamma'}^{1,2}=-t_{\gamma,\gamma'}^{2,1}=-t_{4}$, if $\gamma\ne\gamma'$,
and zero otherwise. To avoid a proliferation of extra parameters in our
analysis, we have decided to fix the values of these hoppings from considerations
previously used in the pnictide context.
Our use of models originally devised for pnictides is simply based on the pragmatic
observation that some pairing tendencies and ferromagnetic regions at large $J$ 
were already observed in those models.\citep{prl2orbdagotto,prb2orbdagotto} It should be
clear though, that our research is mainly motivated by heavy fermion phenomenology.

Following this strategy, then the 
hopping amplitudes $t_{1},$ $t_{2}$, $t_{3}$,
and $t_{4}$  are obtained using the Slater-Koster tight-binding
scheme, and they are given by \citep{Slater-Koster,prl2orbdagotto,prb2orbdagotto}
\begin{eqnarray}
t_{1}&=&-2\left(b^{2}-a^{2}+g^{2}\right)/\Delta_{pd}\nonumber\\
t_{2}&=&-2\left(b^{2}-a^{2}-g^{2}\right)/\Delta_{pd}\nonumber\\
t_{3}&=&-\left(b^{2}+a^{2}-g^{2}\right)/\Delta_{pd}\nonumber\\
t_{4}&=&-\left(ab-g^{2}\right)/\Delta_{pd},
\end{eqnarray}
where the Fe-As hopping amplitudes are $a=0.324(pd\sigma)-0.374(pd\pi)$,
$b=0.324(pd\sigma)+0.123(pd\pi)$, and $g=0.263(pd\sigma)+0.31(pd\pi)$.
$\Delta_{pd}$ is the energy difference between the $p$ and $d$
levels. We have set $(pd\sigma)^{2}/\Delta_{pd}=1$ to fix the
energy scale, and we use $pd\pi/pd\sigma=-0.2$, 
as previously discussed.\citep{prl2orbdagotto,prb2orbdagotto} 
Regarding the couplings $U$,
$J$, and $J'$, note that they are not independent, 
but they are assumed to satisfy the relation $U=U'+2J$, which is strictly valid within a cubic environment for the
full $t_{\rm 2g}$ sector.\cite{re:kanamori63,re:tang98} For an explicit derivation of this relation in the case of
manganites see Ref.~\onlinecite{prdagotto}. 

We have investigated the model defined above using a two-leg ladder of size 
2$\times$$L$, by means of the Density Matrix Renormalization Group 
(DMRG) technique,\cite{re:white92}
under open boundary conditions (OBC), and keeping up to $m$=1400
states per block in the final DMRG sweep. We have carried out $\sim6-13$ sweeps,
and the discarded weight was typically $10^{-6}-10^{-9}$ at the
final sweep. In our DMRG procedure, the center blocks are composed
of 16 states.
We have used a FORTRAN DMRG code to calculate most results, and a C++ DMRG 
code for additional validation.\cite{re:alvarez09}  
We have also confirmed 
some of our results against Lanczos Exact Diagonalization techniques, when possible.

In this work, we will focus on the region of Hubbard and Hund 
parameters where $U_{\rm eff}\equiv U-3J=U'-J<0$,
although a few results for  positive values of $U_{\rm eff}$ will be presented as well.
As shown below, 
in the region where $U_{\rm eff}<0$ there is a robust evidence of 
binding of holes/electrons
close to density $\rho=2$. In this regime $U'<J$. This inequality may not
seem realistic at first sight, since the on-site 
spin triplet formation favored by $J$ (Hund's rules) 
has its origin in the alleviation
of the Coulombic energy penalization caused by $U'$. However, while in the full
five-orbital model for Fe-based compounds $U'<J$ is unphysical for the
reason stated above, the two-orbital model is an {\it effective} model 
and it is
unclear how the main parameters are affected by the projection from five
to two orbitals. Thus, the regime $U'$ comparable to $J$ may not be
unrealistic. Clearly additional investigations are needed to analyze if this
regime of couplings is of relevance for real materials, such as heavy 
fermions. 
Ab-initio calculations are needed
for this purpose (beyond the scope of the present analysis).

\section{RESULTS}

\subsection{Magnetic Properties at $\rho=2$}

\begin{figure}[tb]
\begin{centering}
\includegraphics[scale=0.30]{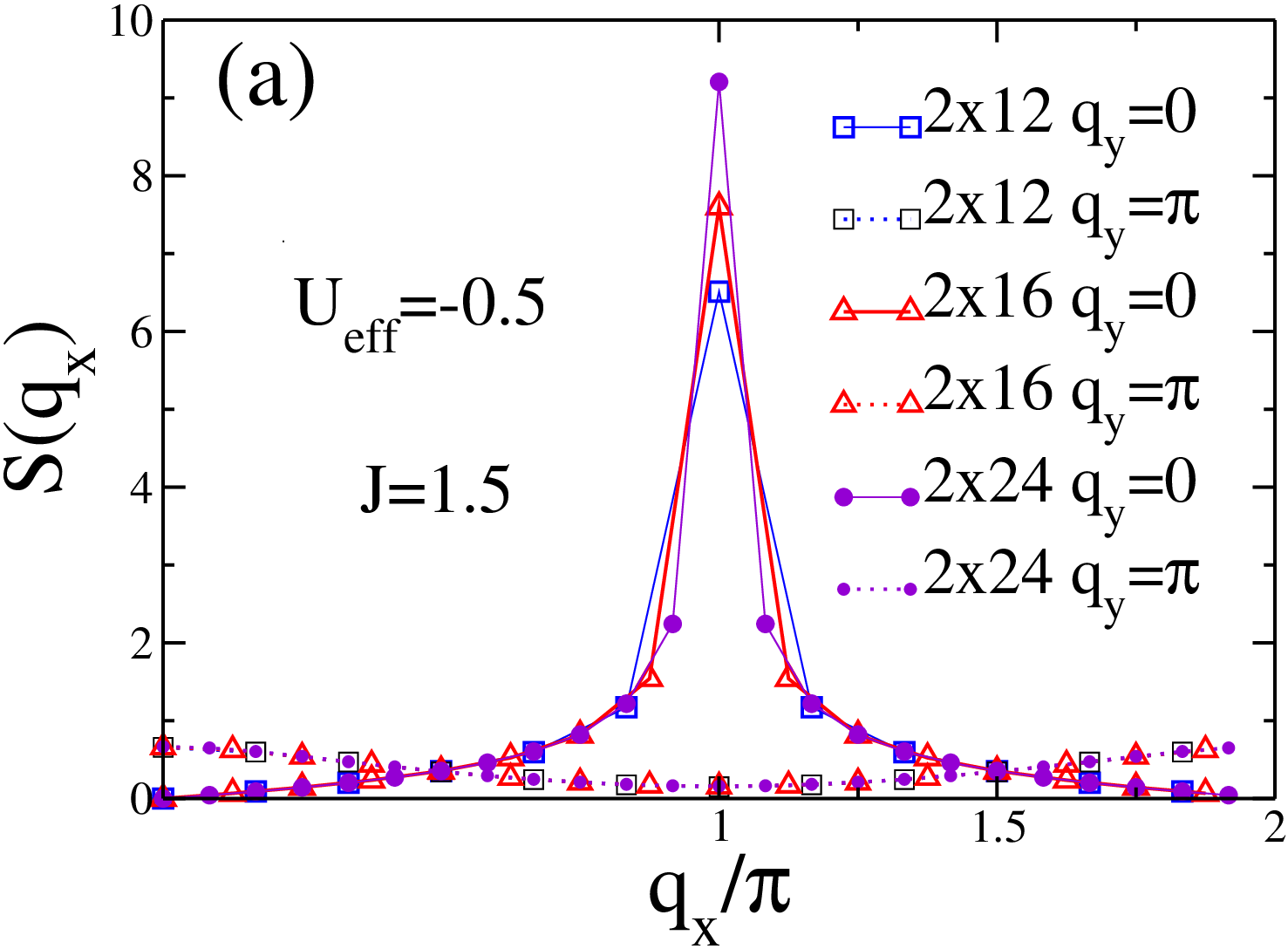} 
\par\end{centering}

\begin{centering}
\includegraphics[scale=0.30]{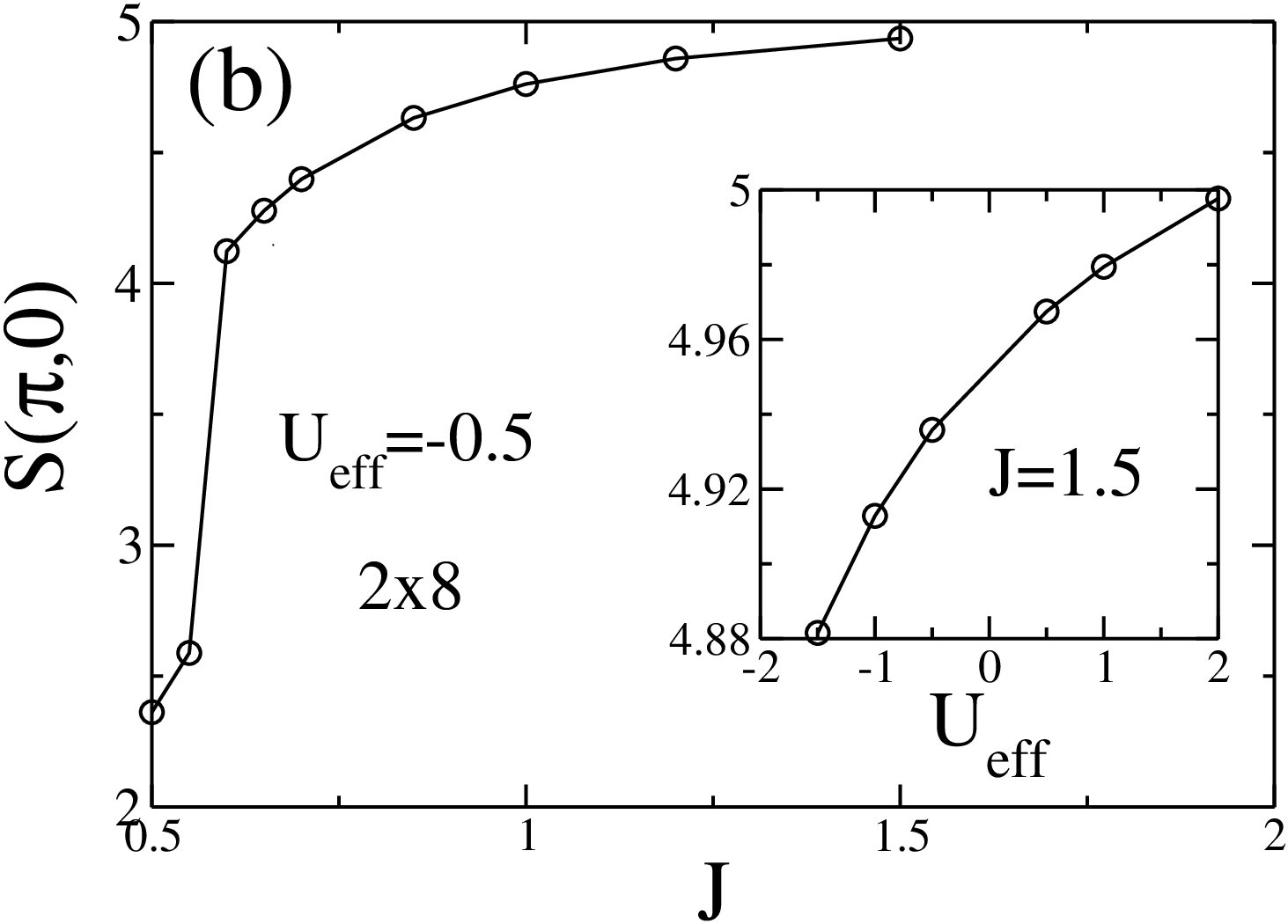}
\par\end{centering}

\begin{centering}
\includegraphics[scale=0.30]{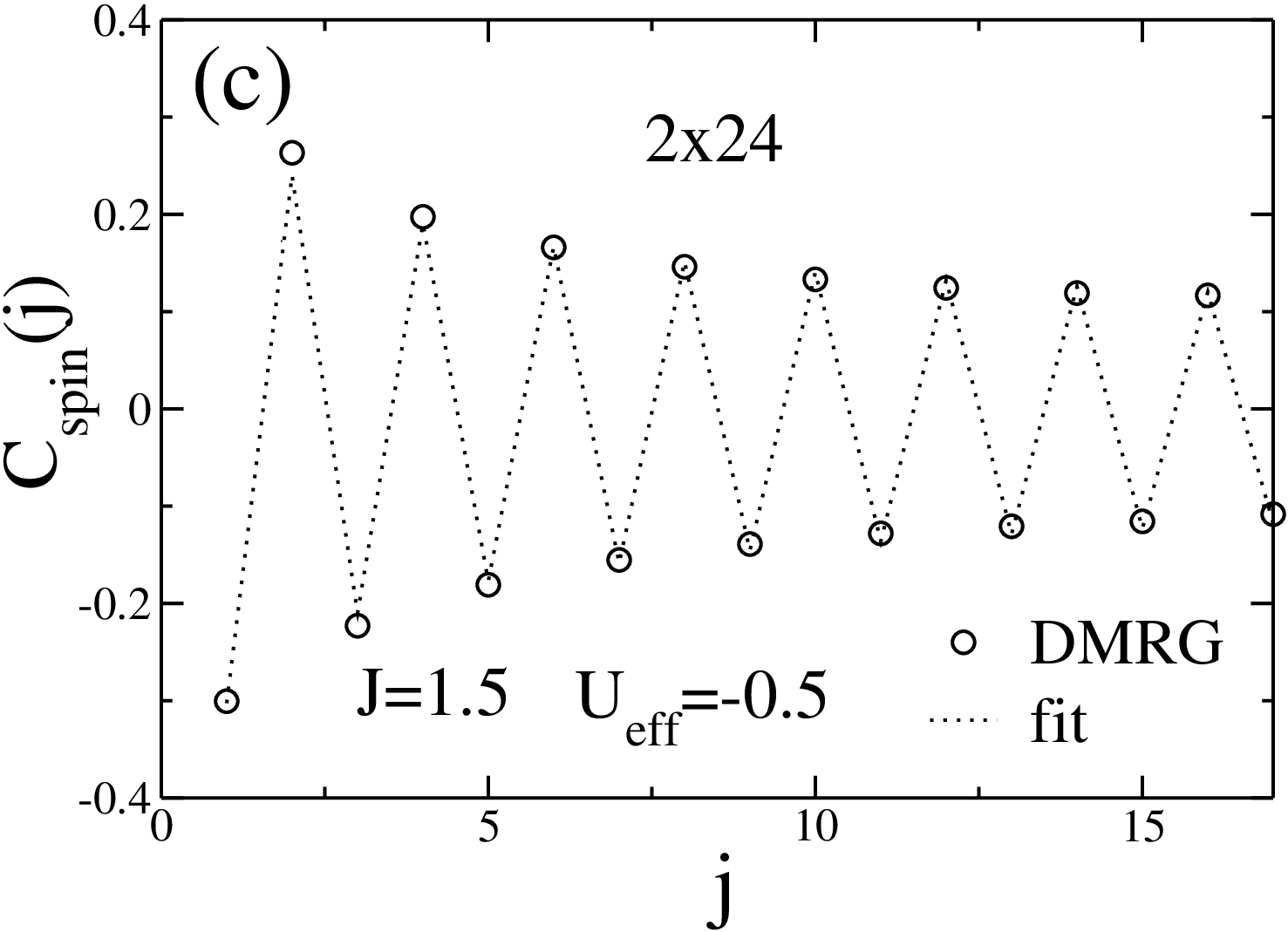}
\par\end{centering}

\caption{\label{fig:1} (Color online). (a) Spin-structure factor $S(q_{\rm x})$
vs. $q_{\rm x}$ for the two-leg ladder system with sizes $L=12,16$, and 24, and for
density $\rho=2$, $J=1.5$, and $U_{\rm eff}=-0.5$. 
(b) $S(\pi,0)$ as a function of $J$ for ladders
with linear sizes $L=8$ and $U_{\rm eff}=-0.5$. The inset shows the magnitude of this peak
as a function of $U_{\rm eff}$, for the coupling $J=1.5$. (c) Spin-spin
correlation $C_{\rm spin}(j)$ vs. $j$ along the long ladder direction, 
for a system with size $L=24$, and
for the couplings $U_{\rm eff}=-0.5$ and $J=1.5$. The dashed curve
is a fit given by Eq.~(\ref{eq:fitting}). }
\end{figure}

Let us start our analysis by focusing on the density $\rho=2$. To investigate
the magnetic order at this density, the Fourier transform of the real-space 
spin-spin correlation function was measured:
\begin{equation}
S\left(\bold{q}\right)=\frac{1}{6L}\sum_{j,j ',\lambda,\lambda '}
e^{iq_x\left(j-j '\right)}
e^{iq_y\left(\lambda-\lambda '\right)}
\left\langle \mathbf{S}_{j,\lambda}\cdot \mathbf{S}_{j ',\lambda '}\right\rangle,
\end{equation}
where
$\mathbf{S}_{j,\lambda}=\mathbf{S}_{j,x,\lambda}+\mathbf{S}_{j,y,\lambda}$.
In Fig.~\ref{fig:1}(a), the spin structure factor $S(\bold{q})$
of the two-leg model is presented for several system sizes. As can be observed in this
figure, there is a robust peak at wavevector $\bold{q}$=$(\pi,0)$, showing the tendency
toward a stripe-like AFM order. This is the analog of the magnetic 
order found in pnictides but
using a two-leg ladder geometry. Similar results, obtained with Exact
Diagonalization on small clusters, were reported before.\citep{prl2orbdagotto,prb2orbdagotto}
Note that the peak increases with the system size, suggesting the development
of a true long-range magnetic order in systems with a higher dimension. 

The results for the spin structure factor show 
that along the $y$($x$)-axes the spins
are aligned following a ferromagnetic (AFM) order, at least at short distances. This 
stripe-like AFM structure 
is present in a wide range of parameters, as shown in Fig.~\ref{fig:1}(b) (inset), 
including
$U_{\rm eff}>0$. 
Neutron scattering measurements for pnictides
also show a similar spin order.\citep{neutronIron}
We have observed that the stripe-like AFM structure appears only when
the  plaquette-diagonal hopping amplitudes ($t_3$ and $t_4$) are of value similar
as those of the nearest-neighbor hopping amplitudes.
If we force $t_3$= $t_4$=0 then the peak in the spin structure factor $S(\bold{q})$ 
appears at wavevector  $\bold{q}$=$(\pi,\pi)$. Note also that for the 
two-leg geometry, the  $(\pi,0)$ AFM state is not, naturally, degenerate 
with the ($0,\pi)$ AFM state. Due to this fact, a study in a ladder geometry 
could make a better connection with the two-dimensional results of the
pnictide materials where $(\pi,0)$ is favored over $(0,\pi)$ 
by a lattice distortion.

For quasi-one-dimensional systems, a true long-range
magnetic order is replaced by a power-law decay of the spin-spin correlations.
Thus, in order to analyze the range of the magnetic order, we have also investigated
the spin-spin correlation function along one of the legs (say, leg 1) defined as
\begin{equation}
C_{\rm spin}(l)=\frac{1}{M}\sum_{|i-j|=l}\left\langle
S_{i,1}^{z}S_{j,1}^{z}\right\rangle ,
\end{equation}
where $M$ is the number of site pairs $(i,j)$ satisfying $l=|i-j|$.
In practice, we have averaged over all pairs of sites separated by distance $l$, in
order to minimize boundary effects (a few sites
at the edges were also discarded while implementing this averaging procedure).

In Fig.~\ref{fig:1}(c), the spin correlation function $C_{\rm spin}(j)$ is shown
for the 2$\times$24 cluster and using $U_{\rm eff}=-0.5$. The dashed line is a fit
of the numerical data with the function 
\begin{equation}
\tilde{C}_{\rm spin}=a\frac{\cos(\pi x)}{x^{1/3}}.\label{eq:fitting}\end{equation}
Similar results were found for the 2$\times$16 cluster. The observed power-law decay
suggests that a two-dimensional system with the same model and parameters would develop 
long-range magnetic order at zero temperature.

\subsection{Doping with two holes or electrons}

Let us now consider the effect of doping with charges this ladder system. 
If tendencies toward the pairing of the extra charges are
unveiled, they would be an indicator that this model could become superconducting
in a two dimensional geometry, for the couplings here considered. 

Let us start with the calculation 
of the binding energy of two doped holes/electrons. This binding energy is
defined as $\Delta_{\rm b}=E(2)+E(0)-2E(1)$, where $E(n)$ is the ground
state energy with $(4L+n)$ holes/electrons ($4L$ is the number of electrons corresponding
to the ``undoped'' limit where there is an electron per orbital and per site). 
On a \emph{finite system} the binding energy
$\Delta_{\rm b}>0$ is positive if the electrons/holes
do not form a bound state,\citep{dagottorev} while in the thermodynamic
limit $\Delta_{\rm b}$ should vanish in the absence of pairing. 
On the other hand, if the extra holes/electrons form a bound state, then $\Delta_{\rm b}<0$
even on a finite cluster, 
and this would be indicative that  effective attractive forces are present in the system.

Before presenting our numerical results for the binding energies, let us 
first consider the following strong coupling regime defined by $-U_{\rm eff}=3J-U=J-U'\gg1$.
As it will be argued below, and as it was discussed in previous investigations
for one-dimensional chains,\citep{1D} in this regime the binding of hole/electrons is clearly present. 

For completeness, let us address this limit in detail, although it is clear that large $J$
compared with $U'$ leads to an effective attractive interaction.\citep{1D}
Let us denote the states of the one-site problem 
via the symbol $\left(\frac{s_{y}}{s_{x}}\right)$,
where the ``arrows'' 
$s_{x}$($s_{y})$ represent the electrons (with their spins projections) at the orbital $x$($y$).
For this one-site problem
with two electrons and the limit considered here where $J$ is large, 
the ground state energy is degenerate. Its 
 value is $e_{2}=U_{\rm eff}$, and
the three corresponding eigenstates are 
$\left(\frac{\downarrow}{\downarrow}\right)$, 
$\left(\frac{\uparrow}{\uparrow}\right)$, and
[$\left(\frac{\uparrow}{\downarrow}\right)$+
$\left(\frac{\downarrow}{\uparrow}\right)$]/$\sqrt{2}$. 
Since $-U_{\rm eff}\gg1$,
we can approximate the Hamiltonian as $H=H_{\rm kinetic}+H_{\rm int}\sim H_{\rm int}$, and the (highly degenerate) ground state energy of the two-leg model
becomes $E(0)=2L \times e_{2}=2L \times U_{\rm eff}$.

Below, the dominant spin arrangement in the ground state at
density $\rho=2$ is shown to guide the discussion 
\[
\left(\begin{array}{ccccc}
\left(\frac{\downarrow}{\downarrow}\right) & \left(\frac{\uparrow}{\uparrow}\right) & \left(\frac{\downarrow}{\downarrow}\right) & \left(\frac{\uparrow}{\uparrow}\right) & \left(\frac{\downarrow}{\downarrow}\right)\\
\left(\frac{\downarrow}{\downarrow}\right) &
\left(\frac{\uparrow}{\uparrow}\right) &
\left(\frac{\downarrow}{\downarrow}\right) &
\left(\frac{\uparrow}{\uparrow}\right) &
\left(\frac{\uparrow}{\uparrow}\right)\end{array}\right),
\]
using only the up and down projections of the spin one states at every site for
simplicity. 
Such a state is to be expected since for $J\gg1$ the alignment of the two spins
at the two orbitals in the same site will occur, and as $U\sim2J\gg1$ then having two electrons
in the same orbital is not allowed. Of course, any other configuration
(such as a fully ferromagnetic state) is also equally likely  as the one shown in the figure 
for $-U_{\rm eff}\gg1$. However,
the hopping terms will lift the large degeneracy, and our numerical results for the
spin-spin correlations presented before indicate that hopping terms
favor the $(\pi,0)$ AFM configuration, for a wide range of couplings in the undoped limit. 
For this reason, here we have chosen to present 
the dominant spin arrangement 
of the ground state as  a stripe-like AFM state, but it could
be fully FM as well.
Regardless of this detail, the arguments we will
use below to obtain the energies $E(n)$ will hold true both for a stripe-like AFM state,
as well as for a FM state. 

Let us now add two extra electrons to the undoped system. In the limit being considered here,
where the hopping amplitudes are negligible, the best way for the system to
minimize its energy is to have the two electrons located on  the same site,
since in this way less on-site ferromagnetic links are broken. The on-site Hubbard $U$ energy
penalization is the same whether the doubly occupied orbitals are at the same site or not, and
since $U'$ is negligible with respect to $J$ in the limit considered, then the 
(effectively attractive) Hund coupling
determines the location of the extra charge, leading to the double 
occupation of both orbitals at the same site.\citep{1D}
In this case, the dominant spin arrangement
of the ground state is

\[
\left(\begin{array}{ccccc}
\left(\frac{\downarrow}{\downarrow}\right) & \left(\frac{\uparrow}{\uparrow}\right) & \left(\frac{\downarrow\uparrow}{\downarrow\uparrow}\right) & \left(\frac{\uparrow}{\uparrow}\right) & \left(\frac{\downarrow}{\downarrow}\right)\\
\left(\frac{\downarrow}{\downarrow}\right) &
\left(\frac{\uparrow}{\uparrow}\right) &
\left(\frac{\downarrow}{\downarrow}\right) &
\left(\frac{\uparrow}{\uparrow}\right) &
\left(\frac{\uparrow}{\uparrow}\right)\end{array}\right).
\]
\noindent Once again, a stripe-like AMF  background is used, since
for $J\sim 1$ and close to the density $\rho=2$ our numerical data show
(see Fig.~\ref{fig:4}(a)) that the $(\pi,0)$ AFM order is the dominant one, but
it could have been FM as well. 

In the limit of couplings 
considered here,
 the ground state energy for the doped two-electron system is
\begin{equation}
E(2)=(2L+5)U_{\rm eff}+8J.
\end{equation}
Note that if the two extra electrons that are in the state 
$\left(\frac{\downarrow\uparrow}{\downarrow\uparrow}\right)$
were to move in opposite directions (after considering the presence of 
small but nonzero hopping terms) 
they would break two
ferromagnetic on-site links. Such a state would have a large energy and it is therefore ``forbidden''
for $-U_{\rm eff}\gg1$. However, if the two extra electrons move ``together''
(i.e. $\frac{\downarrow}{\downarrow}$) in the same direction, forming a spin triplet, then no other
on-site FM links are broken. Thus, to minimize the
energy the two-electrons added to the system must form
a bound state, at least in the limit where the hoppings amplitudes are very small compared
with the Hubbard and Hund couplings
(the same spin dominant picture
works when instead of adding electrons we 
remove two electrons, i.e. for the two-hole problem). 
This argument, known from previous investigations using chains,\citep{1D} explains the
binding of electrons in the limit
$-U_{\rm eff}\gg1$. More explicitly, the binding energy of the doped two-electrons/holes system in the
limit considered here is
given by 
\begin{equation}
\Delta_{b}=-|U_{\rm eff}|.\label{eq:deltab}\end{equation}

Based on the discussion above we conclude that for
$-U_{\rm eff}\gg1$,
there is an indication of pairing, and perhaps superconductivity, in the two-orbital model.
In fact, the on-site interorbital pairing state found here in this extreme regime results to be
a spin triplet and transforms according to the irreducible representation $A_{\rm 2g}$ of the
group $D_{\rm 4h}$.\citep{wan} Interestingly, Exact Diagonalization calculations in two-dimensional
clusters, still in the FM state of the model but $U'>J$, found indications of a pairing state with
the same characteristics and symmetry but with electrons at distance of one lattice spacing from each other.\citep{prl2orbdagotto,prb2orbdagotto}
The argument presented above  for this pairing is actually valid in any dimension,
and also it works for the two doped holes case, as already mentioned. 
To confirm these argumentations,
$\Delta_{\rm b}$ was calculated numerically  for large values of $-U_{\rm eff}$. An excellent
agreement between the numerical data and the analytic expression 
(Eq.~(\ref{eq:deltab})) was found.

\begin{figure}[tb]
\begin{centering}
\includegraphics[scale=0.292]{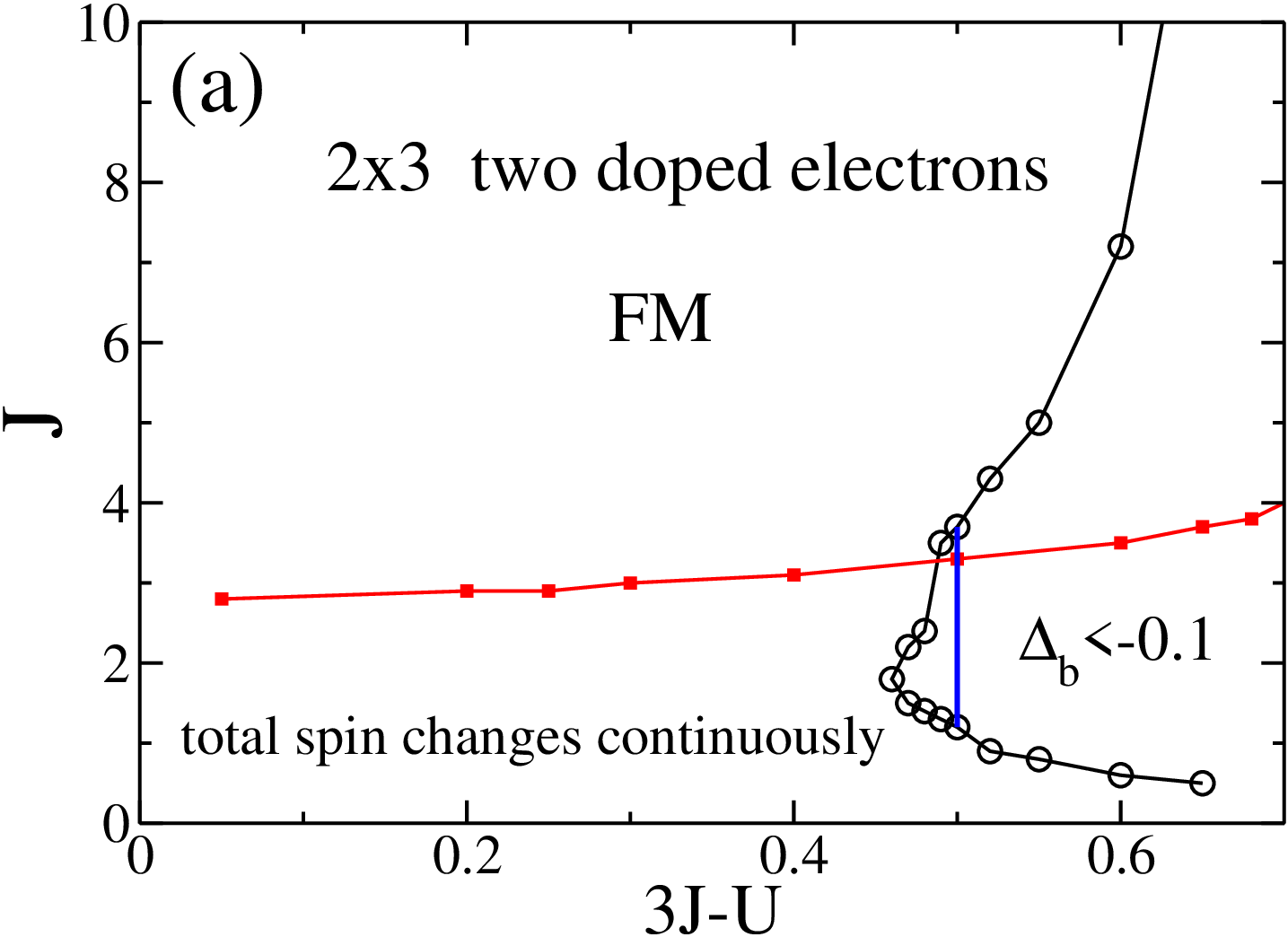}
\par\end{centering}

\vspace*{0.5cm}

\begin{centering}
\includegraphics[scale=0.292]{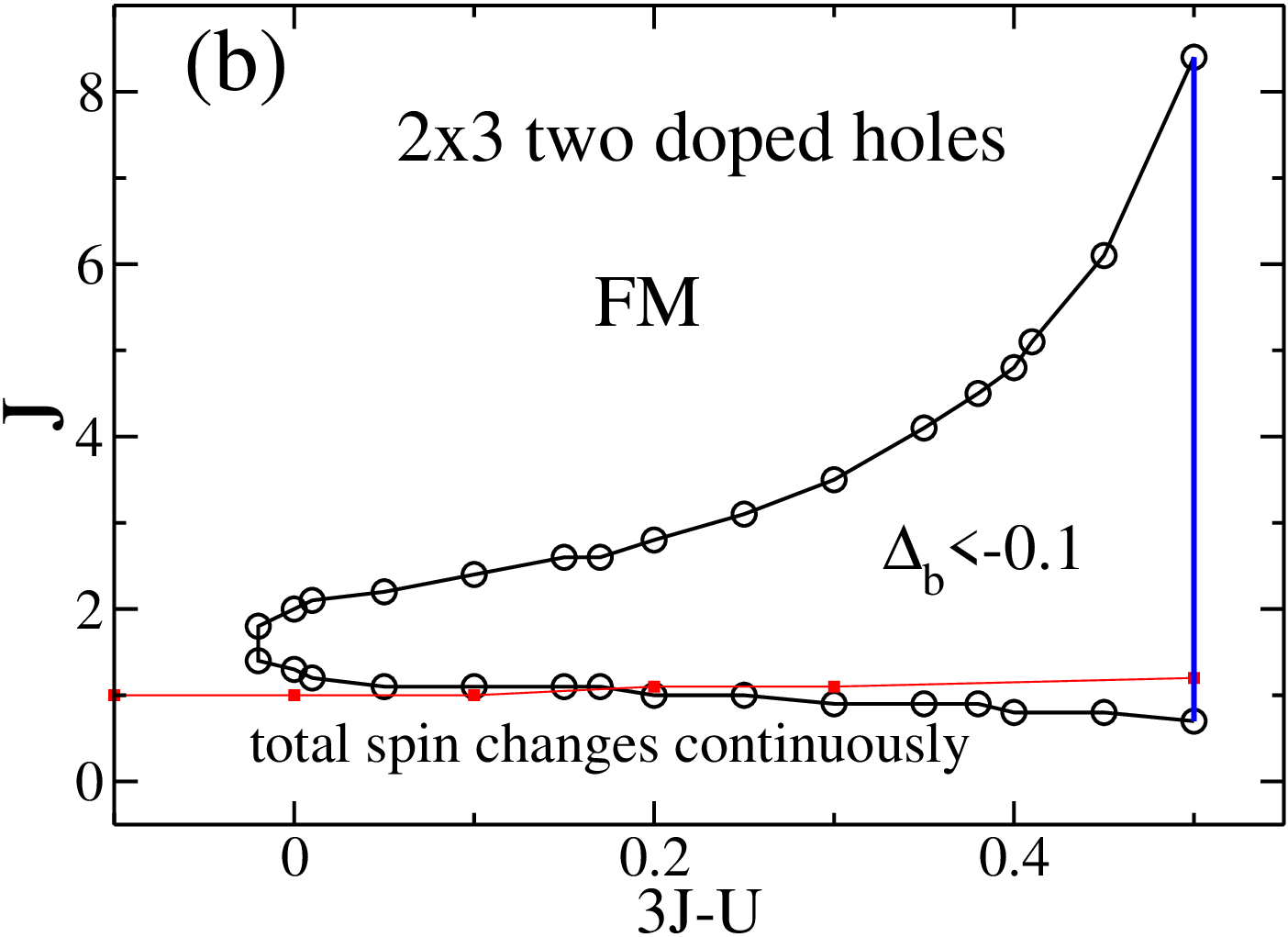}
\par\end{centering}

\vspace*{0.5cm}

\begin{centering}
\includegraphics[scale=0.292]{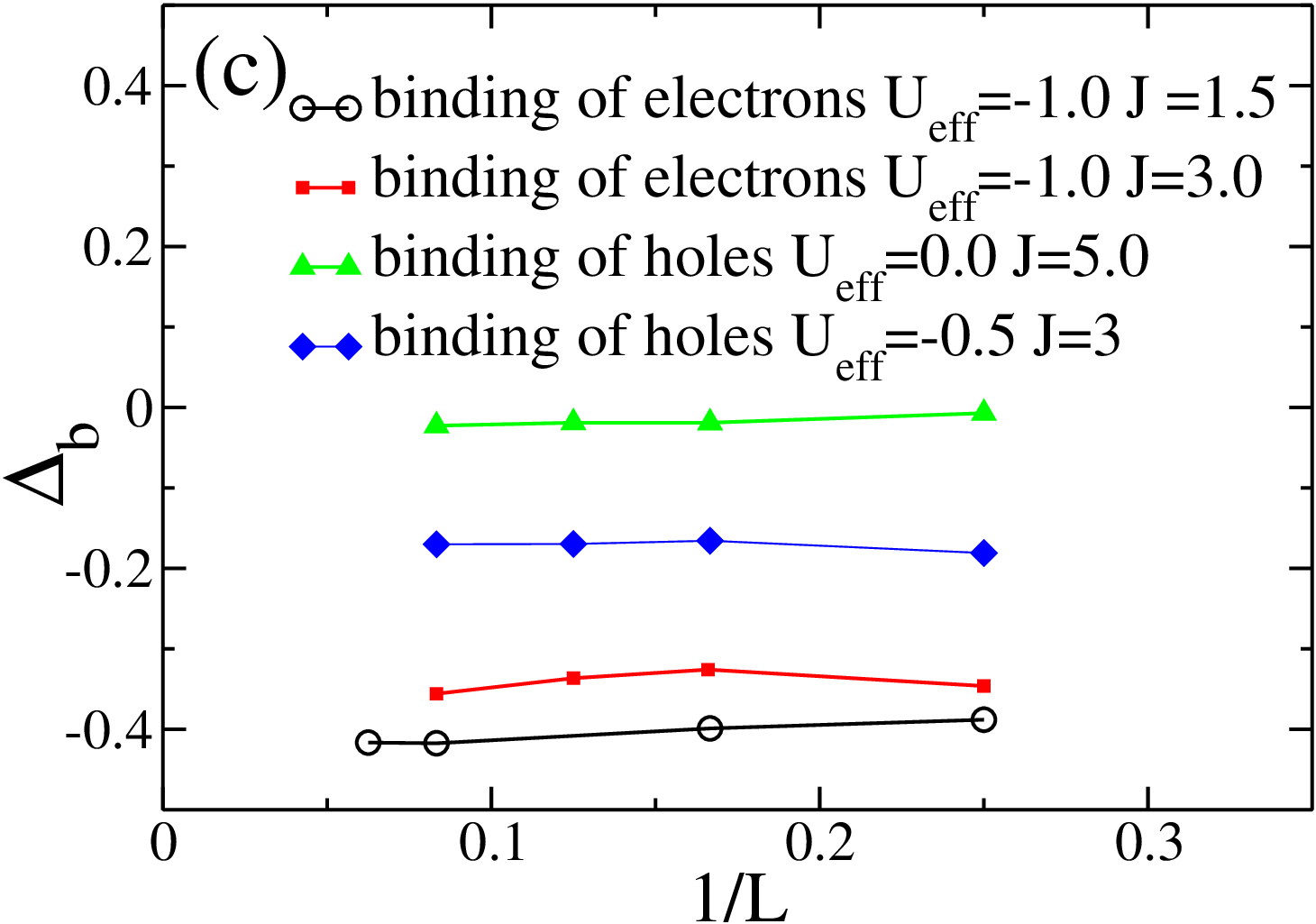}
\par\end{centering}

\caption{\label{fig:2} (Color online). (a)-(b) Phase diagram ($J$ vs. $-U_{\rm eff}=3J-U$)
for the 2$\times$3 cluster 
showing the region where the binding energies of electrons
(a) and holes (b) is smaller than $\Delta_{\rm b}<-0.1$, indicative of pairing. Above the red
line, the total spin saturates to its maximum value. Below this line, the total spin changes
continuously from the maximum value to zero at $J$=0. 
The blue lines show the region of $J$ where $\Delta_{\rm b}<-0.1$ for
$U_{\rm eff}=-0.5$. These lines are the same as presented in Fig.~\ref{fig:3}.
(c) $\Delta_{\rm b}$ vs. $1/L$ for some couplings (see
legend). 
}
\end{figure}

Now the crucial question  is whether there is binding of holes/electrons 
for 
values of $-U_{\rm eff}$ that may be of more relevance for real materials. The
answer to this question appears to be positive.
In fact, the binding of electrons/holes has been observed  numerically for
$U_{\rm eff}\lesssim0$ and several values of the coupling $J$,  as shown in
Fig. 2. \cite{comm}
More specifically, Figs. 2(a)-(b)  show
the region in the $J-(3J-U)$ plane where the binding energies
of electrons (Fig.~\ref{fig:2}(a)) and holes (Fig.~\ref{fig:2}(b)) are less than -0.1,
for the case of a 2$\times$3 cluster. The region
where $\Delta_{\rm b}<-0.1$ was chosen to be represented, 
as opposed to $\Delta_{{\rm b}}$=0, since
previous experience in the context of the cuprates\citep{dagottorev} suggests
that this procedure effectively takes into account size effects better. In practice,
other values for this ``cutoff'' do not alter our qualitative conclusions.
Similar results were found also for the 2$\times$2 cluster.

Thus far, only small clusters have been  considered
because the numerical analysis of large clusters would be too time-consuming, 
particularly with regards to calculating the hundreds of points 
that are required to extract comprehensive
phase diagrams. However, larger system
sizes were considered for a few selected sets of couplings, as shown in
Fig.~\ref{fig:2}(c). 
In this figure, 
$\Delta_{{\rm b}}$ vs. $1/L$ for some couplings is presented. 
Here, it is clearly observed that
in the bulk limit the binding energies of added electrons/holes converge 
to nonzero values
for some coupling sets. Close to $\rho=2$, these results strongly
indicate that there are pairing tendencies
for $U_{\rm eff}\lesssim0$ . Thus, to the extend that future investigations
show that $J$ comparable to $U'$ is a realistic regime for effective two-orbital
models, this provides a possible mechanism for pairing in real materials.

In Figs.~\ref{fig:2}(a)-(b), the magnetic phase diagram for the case of two doped
electrons/holes is also presented. The region above the red (bold) line is a ferromagnetic
phase with the maximum total spin $S_{\rm total}=2L-1$. Below this line, we have observed
that the total spin changes continuously from zero, at $J=0$, up to
its maximum value at $J_{\rm c}$. As observed in these figures and for a large
region of couplings, pairing (and presumably superconductivity) co-exists with ferromagnetic tendencies.

%

\begin{figure}[t]
\begin{centering}
\includegraphics[scale=0.32]{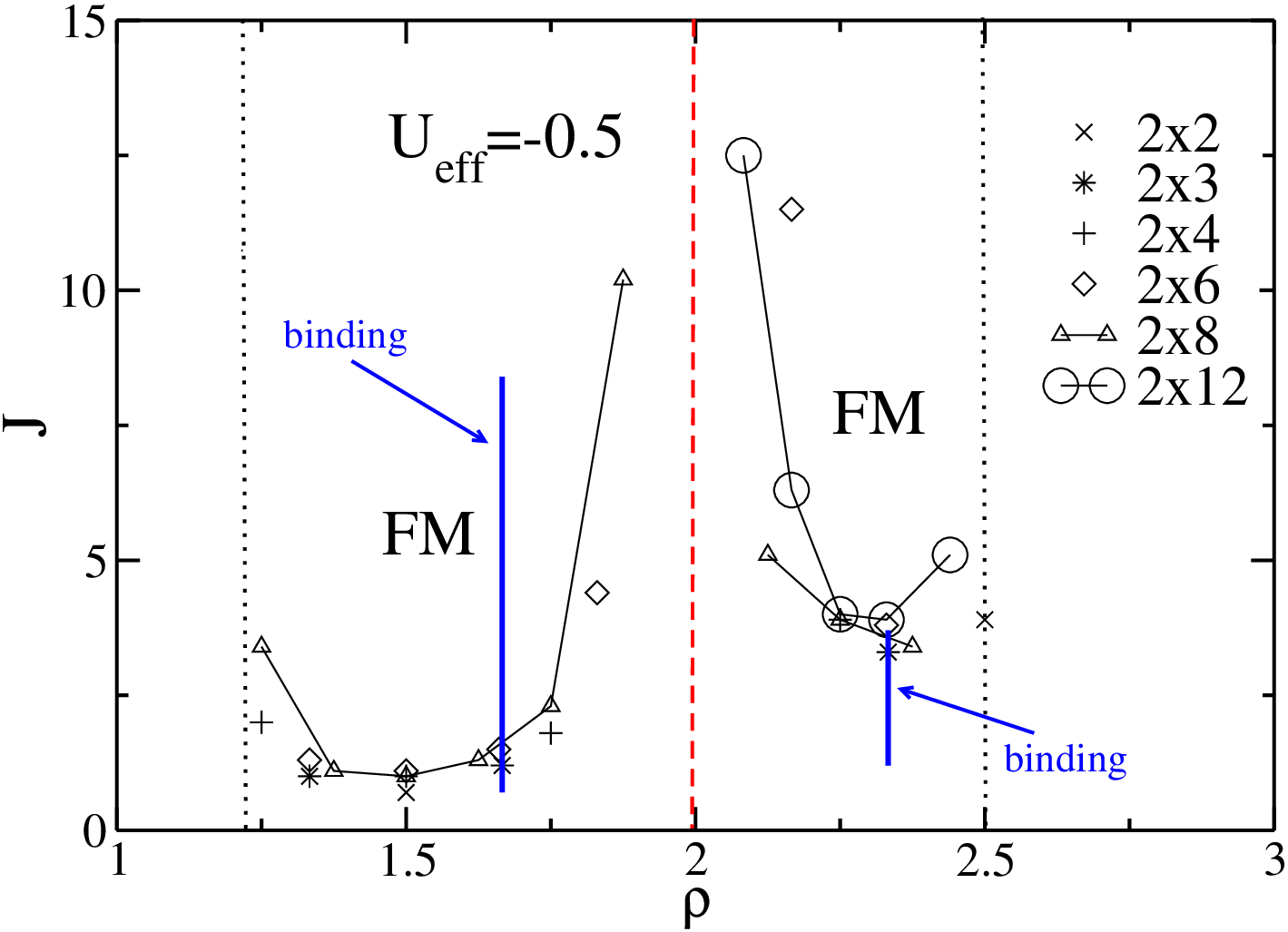}
\par\end{centering}
%
\caption{\label{fig:3} (Color online). (a) Phase diagram of the two-leg
  ladder model Hamiltonian defined in Eq.~1.
The region above the symbols is a fully-saturated ferromagnetic phase (see text). The blue
lines  correspond to regions  where  $\Delta_{\rm b}<-0.1$ for the cluster 2$\times$3
with two doped electrons/holes, as presented
in Figs. 2(a)-(b).
}
\end{figure}

\subsection{Phase Diagram}

In Fig.~\ref{fig:3} the phase diagram ($J$ vs. density)
of the two-leg ladder for $U_{\rm eff}=-0.5$ and several system sizes
(see legend) is presented. For $\rho=2$, it was observed that the total spin  is zero
for $L$ even, and that it can be 0, 1, or 2 for $L$ odd, depending
of the values of $J$. Note that for $L$ odd and $\rho=2$, the $(\pi,0)$ AFM
configuration (+-+-+-) does not have the same number of + and - spins.

For $2<\rho<2.5$ ($1<\rho<2)$, a ferromagnetic phase was found (the
region above the symbols) with magnetic moment per site given by $m=2-\rho/2$
($m=\rho/2)$. The symbols indicate the value of $J_{\rm c}$ where the
total spin  saturates. The critical value $J_{\rm c}$ was determined by
the level crossing of the energies in the sector with $S^{z}=S_{\rm total}^{\rm max}$
and $S^{z}=S_{\rm total}^{\rm max}-1$. Using this procedure, we were able
to obtain $J_{\rm c}$ for large systems. For densities in the ranges $\rho\lesssim1.25$
and $\rho\gtrsim2.5$, we have not found any trace of ferromagnetism. Below the
FM region, it is very hard numerically to determine the total spin  with good accuracy
for large systems. However, our results for the 2$\times$2 and 2$\times$3 clusters 
with two doped electrons/holes suggest that the total spin changes continuously, from maximum
value at $J_{\rm c}$ to zero at $J$=0.
The total spin  can  be extracted from the spin structure factor
at $\bold q=(0,0)$. For a few sets of couplings we also observed, through the value of $S(\bold{q}=(0,0))$, 
that in fact the total spin of the ground state changes continuously 
for the 2$\times$8 cluster as well. 
These results suggest that the total spin varies continuously below the FM
region present in Fig.~\ref{fig:3} for any cluster sizes.  Overall, our results are qualitatively
compatible with those found in one-dimensional systems.\citep{1D}
Note also that tendencies to FM states at robust $J$ were also reported via Exact Diagonalization 
methods on small clusters.\citep{prl2orbdagotto,prb2orbdagotto}

\begin{figure}[t]
\begin{centering}
\includegraphics[scale=0.30]{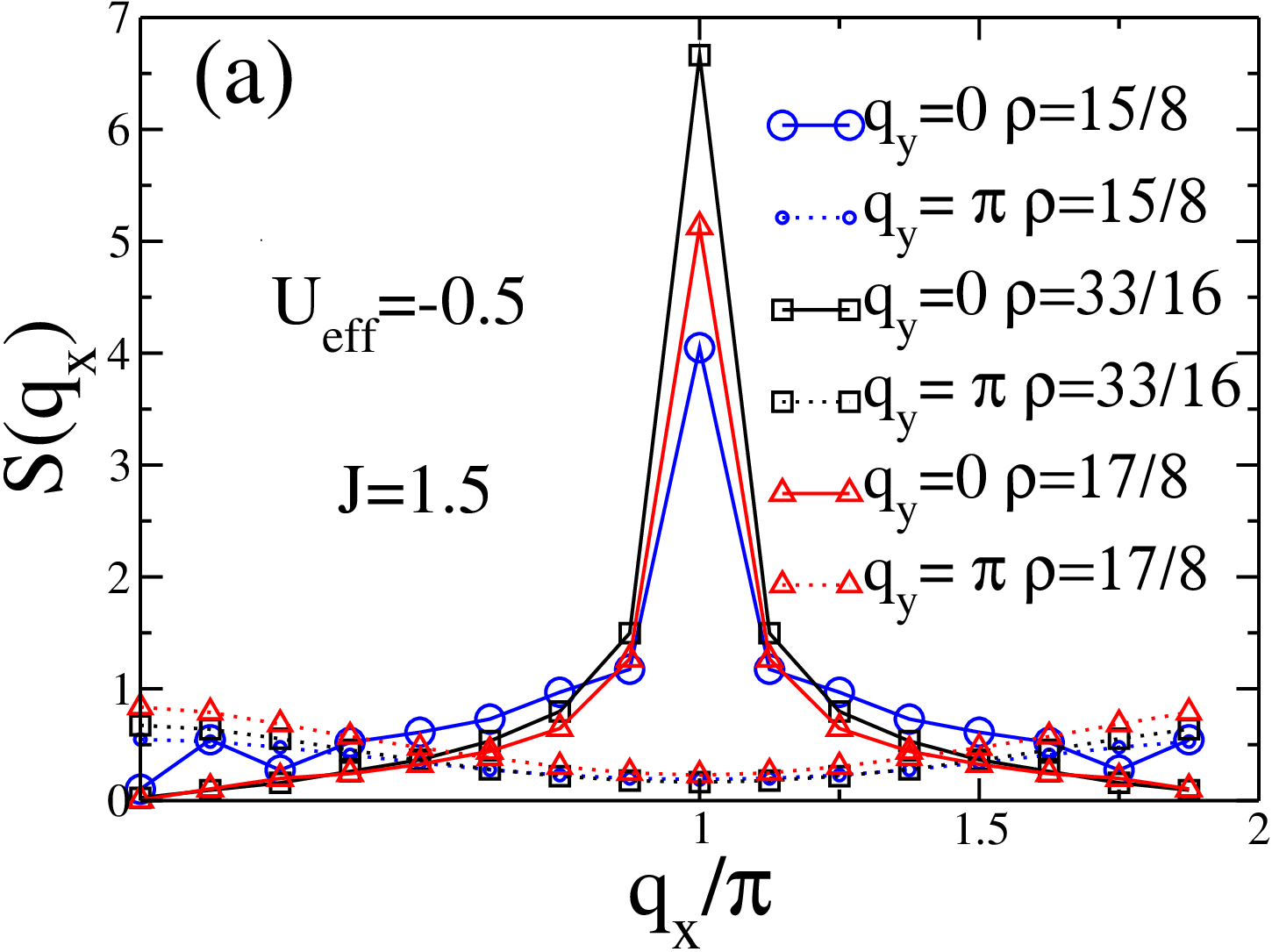}
\par\end{centering}
\begin{centering}
\includegraphics[scale=0.30]{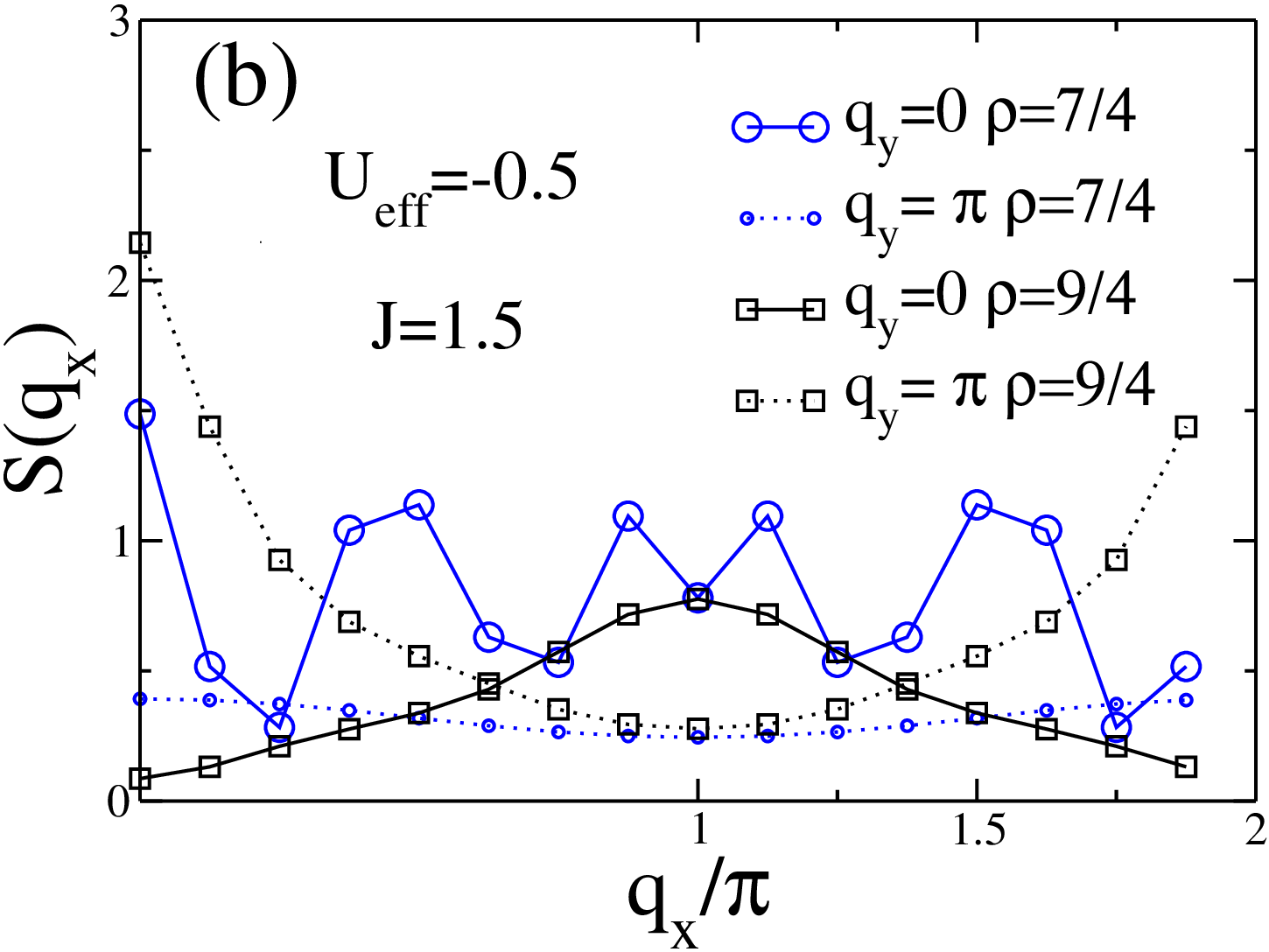}
\par\end{centering}
\caption{\label{fig:4} (Color online). 
Spin-structure factor $S(q_{x})$
vs. $q_{x}$ for the two-leg ladder system with size $L=16$, $J=1.5$, and $U_{\rm eff}=-0.5$.
(a) $S(q_{x})$ for the densities $\rho=15/8$, $\rho=33/16$, and  $\rho=17/8$ (see legend).
(b) $S(q_{x})$ for the densities  $\rho=7/4$ and  $\rho=9/4$.
}
\end{figure}

We believe the FM phase is stabilized by a mechanism that has
the same characteristics as the Double Exchange (DE) mechanism. \cite{zener}
In the  original DE scenario, there are mobile and localized 
degrees of freedom. In the DE mechanism these degrees of freedom are
separated and well defined. Although we do not have localized  
degrees of freedom in our model, from the perspective of one electron 
at a given orbital an electron at the same site but the other orbital
behaves in some respects as a localized spin. For doped systems, 
when an electron moves from one site to the other, 
in order to minimize the kinetic energy and the energy related with the 
Hund coupling, all spins have to be
aligned.

The blue lines in Fig.~\ref{fig:3} are the same that were presented in
Figs.~\ref{fig:2}(a)-(b). We expect that the region of binding extends beyond these
lines up to the density $\rho=2$, forming regions in parameter space where superconductivity
exists inside the phase diagram.

We have also measured the spin structure factor $S(\bold{q})$ away from the undoped density
$\rho=2$. In Fig.~\ref{fig:4}, $S(\bold{q})$ is presented for some particular densities
for the two-leg ladder model with size $L=16$, $J=1.5$, and $U_{\rm eff}=-0.5$. 
As can be observed in Fig.~\ref{fig:4}(a), there is still a peak at
$\bold{q}=(\pi,0)$  for densities close to  $\rho=2$. Note that 
these peaks have smaller intensity than those found for $\rho=2$
in Fig.~\ref{fig:4}(a), for the system with size $L=16$. 
We have also observed that the height of the peak at
$\bold{q}=(\pi,0)$ increases with the system sizes for the densities close to
$\rho=2$. These  results indicate that  a stripe-like AFM magnetic 
order also exists for densities close to $\rho=2$.
As shown in  Fig.~\ref{fig:4}(b), this order does not exist anymore
for $\rho\gtrsim 2.2$ and $\rho\lesssim 1.7$, 
at least within the precision of our calculations, and it is replaced by ferromagnetic
tendencies. 
Note that for the electron doped
case, there is a small peak at $\bold{q}=(0,\pi)$ for densities
$\rho\gtrsim 2.2$.

\section{Conclusion}

Using ladders, we have studied analytically and numerically a two-orbital Hubbard model.
Via the DMRG technique
we were able to investigate the model defined on a two-leg ladder geometry for systems with linear sizes up
to $L=24$. Our spin structure factor data show that for the ``undoped''
density $\rho=2$, a stripe-like AFM order is present, as observed in previous
Exact Diagonalization studies.\citep{prl2orbdagotto,prb2orbdagotto} We have also presented 
evidence for triplet pairing tendencies of added
electrons/holes close to the density $\rho=2$, in some range of couplings, in qualitative agreement with
previous investigations using chains,\citep{1D} and with Exact Diagonalization calculations in a less
extreme FM regime of models for pnictides.\citep{prl2orbdagotto,prb2orbdagotto} More precisely,
we have found that pairing (and presumably superconductivity) and ferromagnetism
co-exist for a large region of parameters in the regime $U'<J$. Even for $U'$ comparable to $J$ our
results still indicate a (mild) tendency to pairing.
Whether this range of couplings for
$U'$ and $J$ is realized in real materials, such as heavy fermions or pnictides, is a matter to be decided via experiments, or with
the help of ab-initio computer simulations.

\begin{acknowledgments}
This research was supported by the Brazilian agencies FAPEMIG and CNPq,
the National Science Foundation 
grant DMR-0706020, the Division of Materials Science and Engineering 
of the U.S. Department of Energy, 
and the Center for Nanophase Materials Sciences, sponsored by the Scientific User
Facilities Division, Basic Energy Sciences, U.S. Department of Energy, under
contract with UT-Battelle. The authors are grateful to Maria
Daghofer for providing us with Exact Diagonalization data to compare our DMRG
results against 
and to Fernando Reboredo and Satoshi Okamoto for useful comments.
\end{acknowledgments}

\begin{thebibliography}{18}
\bibitem{comm} Note that in the analytical discussion of the binding energy,
the plaquette-diagonal hopping amplitudes do not play a significant role.
In fact, we also have found binding of electrons/holes for the cluster 2$\times$4 
at $t_3=t_4=0$ and $U_{\rm eff}<0$.
\end{thebibliography}

\end{document}